\documentclass{article}

\usepackage{arxiv}

\usepackage[utf8]{inputenc} 
\usepackage[T1]{fontenc}    
\usepackage{hyperref}       
\usepackage{url}            
\usepackage{booktabs}       
\usepackage{amsfonts}       
\usepackage{nicefrac}       
\usepackage{microtype}      
\usepackage{lipsum}
\usepackage{authblk}
\usepackage{graphicx}
\usepackage{amsmath}
\usepackage{comment}
\usepackage{xcolor}
\usepackage{adjustbox}
\usepackage{booktabs}
\usepackage{subcaption}
\usepackage{float}
\usepackage{algorithmic}
\usepackage{caption}
\usepackage{cite}
\usepackage{amsmath,amssymb,amsfonts}
\usepackage{algorithmic}
\usepackage{graphicx}
\usepackage{textcomp}
\usepackage{xcolor}
\usepackage{siunitx}
\usepackage{balance}
\usepackage{url}

\title{Simplified Temporal Convolutional-based Channel Estimation for a WiFi Vehicular Communication Channel}

\author[1,2,3]{\textbf{S.~A.~Ngorima}}
\author[1]{\textbf{A.~S.~J.~Helberg}}
\author[1,2,3]{\textbf{M.~H.~Davel}}

\affil[1]{Faculty of Engineering, North-West University, South Africa}
\affil[2]{Centre for Artificial Intelligence Research, South Africa}
\affil[3]{National Institute for Theoretical and Computational Sciences, South Africa}
\affil[ ]{\texttt{aldringorima@gmail.com}}
\date{}
\date{\vspace{-2em}}
\begin{document}
\maketitle 
\begingroup
\renewcommand{\thefootnote}{}
\footnote{\copyright~2025 IEEE. Personal use of this material is permitted. Permission from IEEE must be obtained for all other uses, in any current or future media, including reprinting/republishing this material for advertising or promotional purposes, creating new collective works, for resale or redistribution to servers or lists, or reuse of any copyrighted component of this work in other works. The Version of Record is available online at: \url{https://doi.org/10.1109/WAC63911.2025.10992609}.}
\endgroup

\begin{abstract}
Channel estimation in vehicular communication is a crucial element in the advancement of intelligent transportation systems. However, the use of pilot signals in the IEEE 802.11p standard is insufficient for accurate channel estimation in high-mobility scenarios. Data pilot-aided (DPA) estimation helps address this, but suffers from demapping errors. We propose a simplified Temporal Convolutional Network-based estimator (DPA-TCN) trained on a mixed signal-to-noise ratio dataset to improve estimation performance and reduce computational complexity. Our DPA-TCN estimator achieves a bit error rate comparable to a state-of-the-art long-short-term memory network with DPA and temporal averaging (LSTM-DPA-TA) while reducing the complexity of the model by approximately 65\%.  

\end{abstract}

\section{Introduction}
\label{introdcution}
In intelligent transportation systems (ITS), ensuring reliable and efficient vehicular communication is crucial for enhancing overall transportation safety and performance. Accurate and timely data transmission between vehicles and infrastructure is heavily dependent on precise channel estimation. However, this task is challenging due to the dynamic nature of vehicular channels. The channel in these networks is considered doubly selective, indicating that the channels vary in time and frequency. These variations are caused by two main factors: multipath propagation, where signals reflect off various surfaces, and the high mobility of vehicles, leading to rapid changes in the communication environment. Together, these factors complicate the process of accurate channel estimation, making it a significant challenge in the development of ITS. The IEEE 802.11p standard~\cite{4526014}, approved and specified for vehicular communication to support ITS, faces limitations in scenarios that have high mobility. Specifically, the number of pilot subcarriers allocated in this standard tends to be insufficient, and relying solely on these pilots has proven to be less effective~\cite{gizzini2020deep}.


Increasing the number of pilots for transmission reduces spectral efficiency~\cite{pan2021channel}. To address the problem of insufficient pilots, a method known as data pilot-aided (DPA) estimation is commonly used~\cite{7421323}. This technique involves using data subcarriers for channel estimation. However, the presence of noise and multipath effects causes demapping errors which make it difficult to obtain accurate data pilots using only DPA as an estimation method~\cite{han2019deep}.
In recent years, researchers have suggested using estimators that use deep learning (DL) to capture important features in vehicular channels~\cite{gizzini2020joint, gizzini2020deep, pan2021channel, 10279602}. These estimators are based on the DPA method. 
Among these DL estimators, LSTM-DPA-TA~\cite{gizzini2021temporal} is one of the state-of-the-art (SOTA) in terms of bit error rate (BER) performance. 
However, the computational complexity of this estimator restricts the adaptation of this estimator in real-world applications.


In this work, we propose a simplified Temporal Convolutional Network (TCN) estimator that takes DPA estimation as input for the estimation of vehicular channels. Our aim is to address both the computational and BER performance challenges in vehicular channels by leveraging the TCN's parallel nature and one-dimensional convolutional operations. Furthermore, we train the TCN estimator on a mixed SNR dataset that incorporates various channel variations based on results from~\cite{NgorimaSACAIR2024}, rather than using a dataset generated solely at high SNR as in previous studies~\cite{gizzini2020deep, ngorimaSA, gizzini2020joint, gizzini2021temporal, pan2021channel}. The DPA process solves the insufficient pilot problem, while the TCN uses the DPA estimates to accurately estimate the channel state. We evaluate our proposed DPA-TCN estimator on a realistic vehicle-to-vehicle same direction with wall (VTV-SDWW) tapped delay line (TDL) vehicular channel model specified in~\cite{4350097}. Performance assessment is based on BER, Normalised Mean Square Error (NMSE) and complexity metrics.


\section{System Model}

This section provides an introduction to the IEEE 802.11p standard specification and discusses various channel estimation techniques following~\cite{Ngorima2024}. We will cover the widely used least squares (LS) method, which is considered a baseline in our work. Furthermore, we will dive into the DPA method, commonly utilised for initial channel estimation in various estimators, to highlight the importance of DL post-processing. Lastly, we will compare the performance of these methods with the SOTA, LSTM-DPA-TA.


The IEEE 802.11p standard operates in the frequency band ranging from 5.0 GHz to 5.9 GHz, utilising a 10 MHz bandwidth \cite{4350097}. It employs Orthogonal Frequency Division Multiplexing (OFDM) with 64 subcarriers for data transmission.
Of these 64 subcarriers, 52 are used for data and pilots, while the remaining 12 at indices \{\num{-32},\dots,\num{-27},0,27,\dots,31\} serve as guard bands and DC offsets. Specifically, four pilot subcarriers are located at indices \{\num{-21},\num{-7}, 7, 21\}, with the remaining 48 subcarriers designated for data transmission. 


The IEEE 802.11p frame structure includes a preamble for signal detection, timing synchronisation, and channel estimation, followed by a signal field for transmission parameters. This work assumes perfect synchronisation and considers a frame structure with two extended preambles at the start, followed by \textit{I} OFDM data symbols.

Given this structure, the signal received on the subcarrier $[k]$ at time $i$ can be expressed as
\begin{equation*} Y_{i}[k]=H_{i}[k]X_{i}[k]+N_{i}[k],\tag{1} \label{rec_signal}\end{equation*}
where $X_i[k]$, $H_i[k]$ and $N_{i}[k]$ are the transmitted symbol, the channel frequency response and the Gaussian noise, respectively. 

In this work, we investigate the VTV-SDWW vehicular channel model~\cite{4350097}. This model simulates communication between two vehicles moving in the same direction, separated by a central wall, at a distance of 300-400 metres. In this scenario, vehicles travel at 100 km/h with a Doppler shift of $f_d$ = 550 Hz, using 16QAM modulation.

\subsection{Least Squares (LS) Estimation}

The LS estimation approach is a basic method commonly used in wireless communications. 
The LS approach uses two adjacent received symbols, as shown below:

\begin{equation*} \hat { {H} }_{\text {LS}}[k] = \frac { {Y} ^{(p)}_{1}[k] + {Y} ^{(p)}_{2}[k]}{2 {p} [k]}, \tag{2}\label{eqn:ls}\end{equation*}

where $p[k]$ is the known preamble, ${Y} ^{(p)}_{1}[k]$ and ${Y} ^{(p)}_{2}[k]$ are the preambles received in the $[k]$ subcarrier.

\subsection{DPA}
\label{DPA}
DPA addresses the pilot limitation issue in vehicular communications, without altering the IEEE 802.11p standard frame structure by introducing data pilots. As a result, DPA is recognised as the primary channel estimation method in vehicular communications~\cite{gizzini2020deep}. 
First, DPA obtains an initial estimate of the frame using the LS estimate at the preambles (as shown in (\ref{eqn:ls})).

 
Then, the equalised symbols $Y_{i}^{\text{eq}}[k]$ are obtained for the subsequent symbols by using the previous estimate to equalise the current symbol as follows:

\begin{equation*} Y_{i}^{\text{eq}}[k]= \frac{Y_{i}[k]}{\hat{H}_{i-1}^{\text{DPA}}[k]},\hat{H}_{0}^{\text{DPA}}[k]=\hat{H}_{\text {LS}}[k].\tag{3}\label{eqn:yeq}\end{equation*}

The equalised data symbol $y_{i}^{eq}[k]$ is remapped to the nearest constellation point $d_i[k]$ using the remapping operation $\mathfrak{R}\left(\cdot\right)$ 

\begin{equation*}d_{i}[k]=\mathfrak{R}\left(\frac{Y_{i}[k]}{ \hat{H}_{i-1}^{\text{DPA}}[k]}\right), \hat{H}_{0}^{\text{DPA}}[k]=\hat{H}_{\text {LS}}[k],\tag{4}\label{eqn:conse}\end{equation*}  

%
The DPA channel estimate is then updated as follows:

\begin{equation*} \hat{H}_{i}^{\text{DPA}}[k]=\frac{Y_{i}[k]}{d_{i}[k]}.\tag{5}\label{eqn:fina_DPA}\end{equation*}

\subsection{LSTM-DPA-TA} 
The LSTM in this estimator uses the received OFDM symbol as input, performs the DPA process on the LSTM output, and then applies a time-average filter. The LSTM-DPA-TA method was developed in~\cite{gizzini2021temporal} and was evaluated on several channels, including the VTV-SDWW channel used in this work. We are using the LSTM-DPA-TA estimator for comparison purposes with our proposed DPA-TCN estimator.

\section{Proposed DPA-TCN Estimator}
\label{TCN}
This section outlines the architecture of the DPA-TCN estimator proposed in this work.

\subsection{Temporal Convolutional Networks (TCNs)}

TCNs are designed to handle sequential data by leveraging techniques from Convolutional Neural Networks (CNNs), as introduced by LeCun~\cite{lecun1995convolutional}. TCNs use causal and dilated convolutions along with residual connections to capture sequential dependencies. This is done to capture temporal dependencies within the input data using one-dimensional kernels.

\begin{figure}[htbp]
    \centering
    \includegraphics[width=0.38\textwidth]{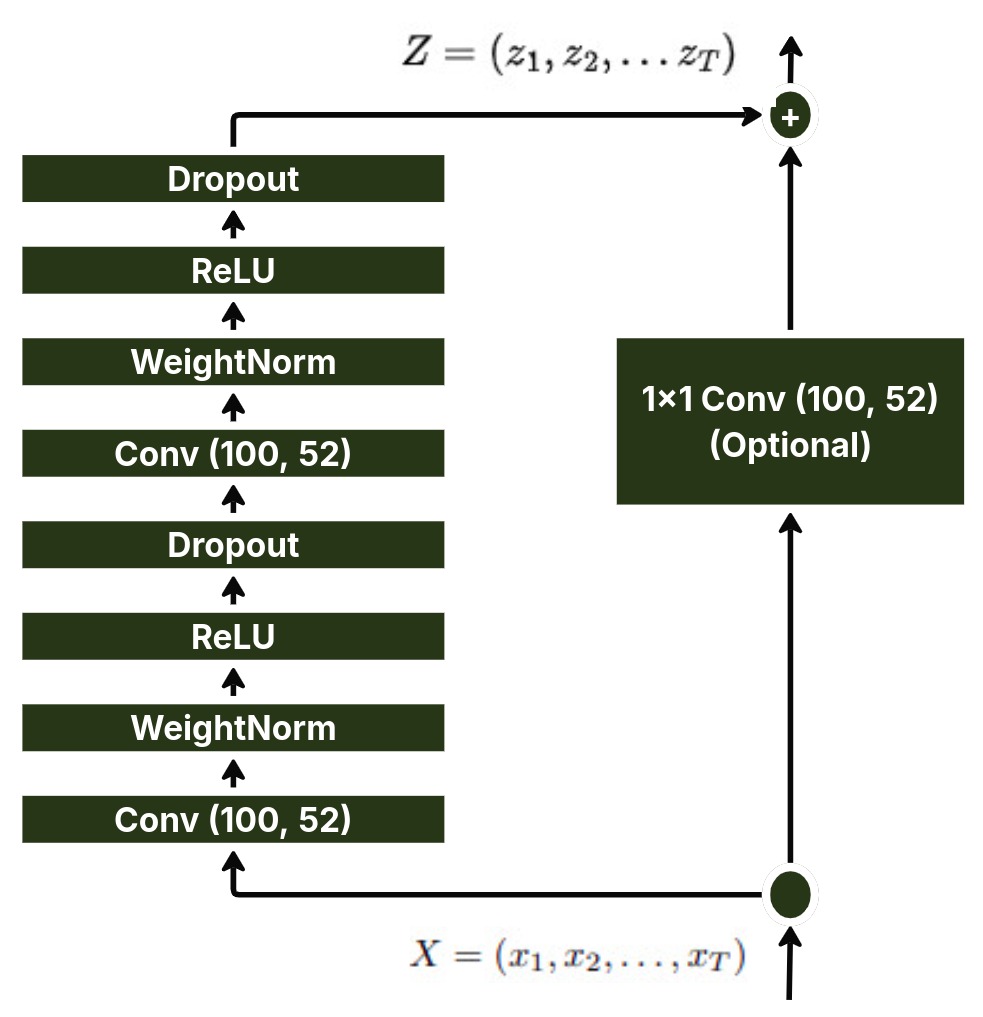}
    \caption{TCN Residual block where 100 are the interleaved real and imaginary parts considered as features and 52 are complex active subcarriers which are considered as timesteps~\cite{bai2018empirical}}
    \label{fig:RB}
\end{figure}

The goal is to predict the outputs ${z_0, \ldots, z_T}$ based on the input sequence ${x_0, \ldots, x_T}$. To predict $z_t$ at time $t$, only the inputs ${x_0, \ldots, x_t}$ are used. To keep the input and output dimensions consistent, TCNs apply padding of length (kernel size - 1) at the start of the frame or the input sequence. Causal convolutions avoid future leakage by only convolving an output at time \textit{t} with elements from time \textit{t} or earlier. An optional residual connection allows the input to be added to the output if the output has dimensions different from the input~\cite{bai2018empirical}.


Furthermore, the residual block typically includes dilated causal convolution layers, nonlinearity layers, and dropout layers after each convolutional layer to effectively prevent overfitting and enhance the model's generalisation capabilities.


\subsection{Dataset Generation}



The dataset for training the proposed DPA-TCN comprises 18,000 time specific frames with SNR levels ranging from 0 dB to 40 dB in 5 dB steps. We generate 2,000 frames at each SNR level, each containing transmitted and received versions of 50 OFDM symbols. Each OFDM symbol has complex values for 52 active subcarriers (48 data and 4 pilot subcarriers). The 16QAM symbols received are separated and interleaved per subcarrier and time slot, resulting in an input matrix of real values with dimensions $52 \times 100$, and an output matrix (excluding pilot signals) of $48 \times 100$. These samples are batched for processing. 

The dataset for training the LSTM-DPA-TA and TCN-DPA estimator is generated as above at a fixed SNR of 40dB.

For all the estimator models, 12,000 frames are used for training, 4,000 frames for validation, and an independent set of 2,000 frames for testing.

Note that in this work, we treat subcarriers as time steps. The input of the TCN is a 3D tensor of shape $(N, C, T)$, where $N$ is the batch size, $C$ is the number of features (100 features) and $T$ is the number of time steps (52 subcarriers). For a batch size of 128, the input tensor shape is $(128, 100, 52)$. A one-dimensional kernel size of 2 processes two consecutive time steps across the input sequence.



\subsection{TCN estimator}

This section describes the proposed DPA-TCN estimator where the input is a DPA-estimated frame $\hat{H}_{i}^{\text{DPA}}[k]$ which consists of $\hat{H}_{i}^{D}$ (estimated data subcarriers) and $\hat{H}_{i}^{P}$ (estimated pilot subcarriers) obtained following (\ref{eqn:ls}) to (\ref{eqn:fina_DPA}). 

\begin{figure}[htbp]
    \centering
    \includegraphics[width=0.85\linewidth]{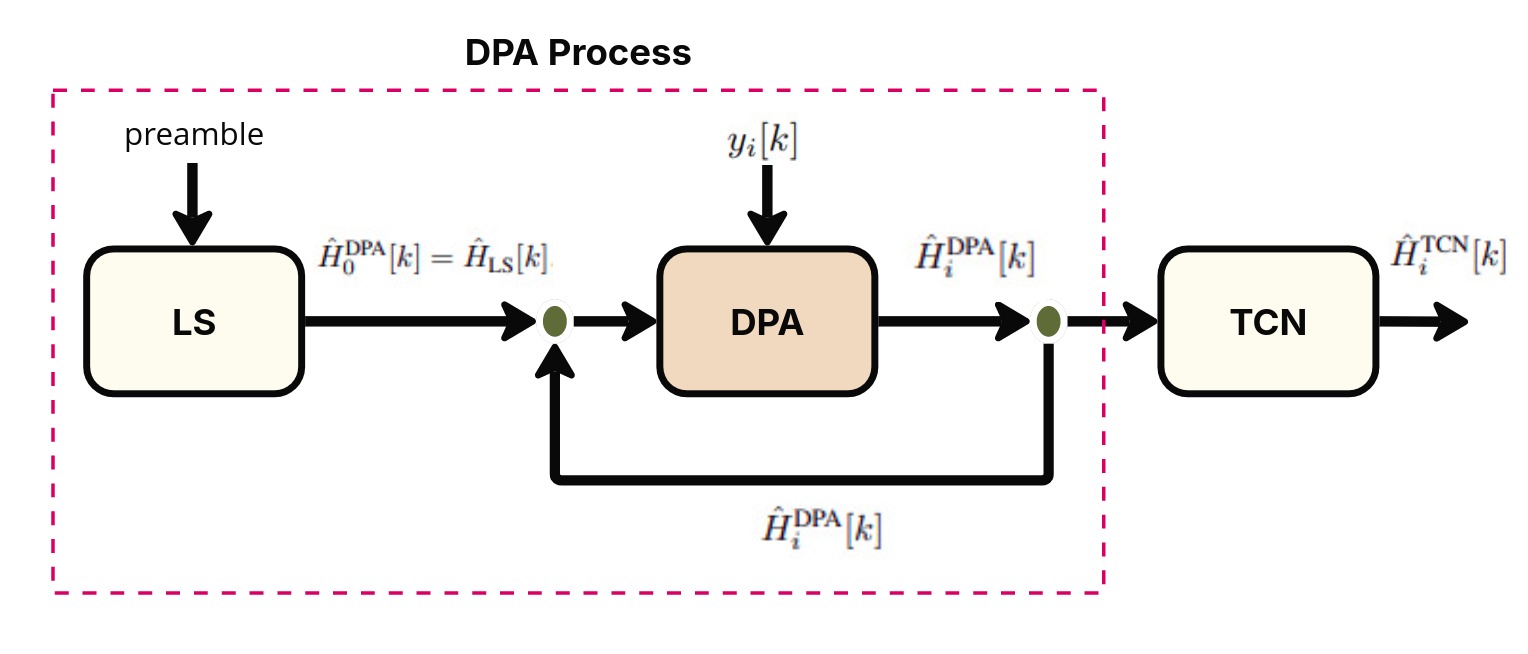}
    \caption{DPA-TCN estimator process}
    \label{fig:error_comp}
\end{figure}

The proposed DPA-TCN estimator is shown in Fig.~\ref{fig:error_comp}.
The final channel estimate \(\hat{H}_{i}^{\text{TCN}}[k]\) is updated based on (\ref{eqn:final_TCN_est}): 
%
\begin{equation*}
\hat{H}_{i}^{\text{TCN}}[k] = f_{\text{TCN}}(\hat{H}_{i-1}^{D}, \hat{H}_{i-1}^{P}. W),\tag{6}\label{eqn:final_TCN_est}\end{equation*}

Here, \( f_{\text{TCN}} \) is a TCN function that processes the input estimates \(\hat{H}_{i-1}^{D}\) and \(\hat{H}_{i-1}^{P}\) to produce the refined channel estimate \(\hat{H}_{i}^{\text{TCN}}[k]\). The function \( f_{\text{TCN}} \) is parameterised by the weight matrix \( W \), which includes the learnable filters and biases of the TCN's convolutional layers. These weights are optimised during training to minimise the estimation error.


In Fig.~\ref{fig:error_comp}, the LS block initialises the channel estimation using the preamble, producing $\hat{H}_{\text{LS}}[k]$. The DPA block iteratively refines the estimation by utilising both the received signal $y_i[k]$ and the prior channel estimate $\hat{H}^\text{DPA}_i[k]$, resulting in $\hat{H}^\text{DPA}_i[k]$. The final TCN block further refines the estimate to obtain the final channel estimate $\hat{H}^\text{TCN}_i[k]$. Arrows indicate the flow of data and the iterative feedback loop within the DPA block.

\subsubsection{Hyperparameter Tuning}
\label{sec:hyper}
We use Bayesian optimisation (BO) to find the best hyperparameter combination for model performance. BO is an iterative algorithm that uses a probabilistic model, commonly a Gaussian process, to approximate an unknown objective function. 
For the TCN estimator, we optimise the learning rate, number of layers, the size of the kernel, the dropout, and the StepLR parameters (step size and gamma) using Weights \& Biases (WandB)\footnote{\url{https://wandb.ai/}} and Optuna \cite{akiba2019optuna}, a BO-backed hyperparameter tuning tool. 
The results of the experiment are recorded and tracked by WandB.



We conduct 50 trials, each running for 200 epochs, to comprehensively explore the hyperparameter space and observe the effects of hyperparameter changes. Convergence is determined using the validation loss. The hyperparameter search space and the best combination are shown in Table \ref{tab:hyperparameters}. 

\begin{table}[htbp]
    \begin{center}
    \renewcommand{\arraystretch}{1.5}
    \caption{Best TCN Hyperparameters}
    \begin{tabular}{|c|c|c|}
    \hline
    \textbf{Hyperparameter} & \textbf{Search Space} & \textbf{Best Value} \\
    \hline
    Learning Rate & $1 \times 10^{-5}$ to $1 \times 10^{-2}$ & 0.0006 \\
    \hline
    Number of Layers & 1 to 5 & 4 \\
    \hline
    Kernel Size & 2 to 5 & 2 \\
    \hline
    Dropout & $10^{-5}$ to 0.5 & 0.17 \\
    \hline
    StepLR Step Size & 10 to 50 & 21 \\
    \hline
    StepLR Gamma & 0.5 to 1 & 0.9 \\
    \hline
    Epochs & 0 to 200 & 156 \\
    \hline
    \end{tabular}
    \label{tab:hyperparameters}
    \end{center}
\end{table}
%
%
\section{Analysis and Results}
\label{results}
In this section, the proposed DPA-TCN estimator is compared to other estimators namely the TCN-DPA, LSTM-DPA-TA, LS, DPA estimators as well as the ideal case with perfect channel knowledge. Furthermore, we present a comparative analysis of the computational complexity of these estimators. 

Firstly, we perform a hyperparameter search for the LSTM model in the LSTM-DPA-TA estimator on the dataset generated at 40 dB using Optuna and WandB, following the same experimental setup discussed in Section~\ref{sec:hyper}. The following best parameters are obtained for the LSTM model: a learning rate of 0.003, LSTM size = 128, StepLR step size = 12 and StepLR step gamma = 0.9. Then, this LSTM architecture is combined with DPA and TA as specified in~\cite{gizzini2021temporal} for comparison with the proposed DPA-TCN estimator. 

\subsection{BER}
\begin{figure}[htbp]
    \centering
    \begin{subfigure}[b]{0.49\linewidth} 
        \includegraphics[width=\linewidth]{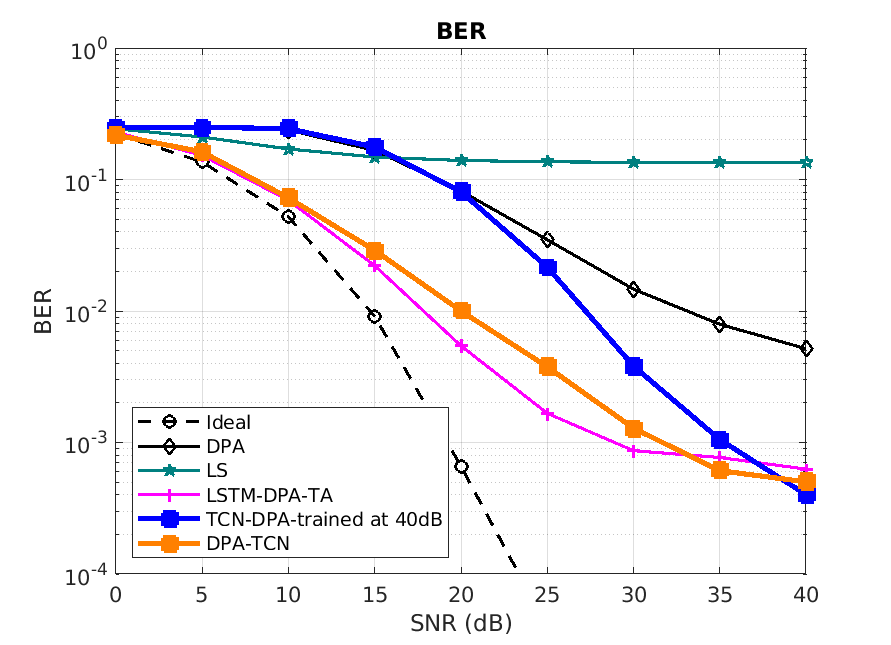} 
        \caption{BER performance of various estimators.}
        \label{fig:TCN-no-TA-BER}
    \end{subfigure}
    \hfill 
    \begin{subfigure}[b]{0.48\linewidth} 
        \includegraphics[width=\linewidth]{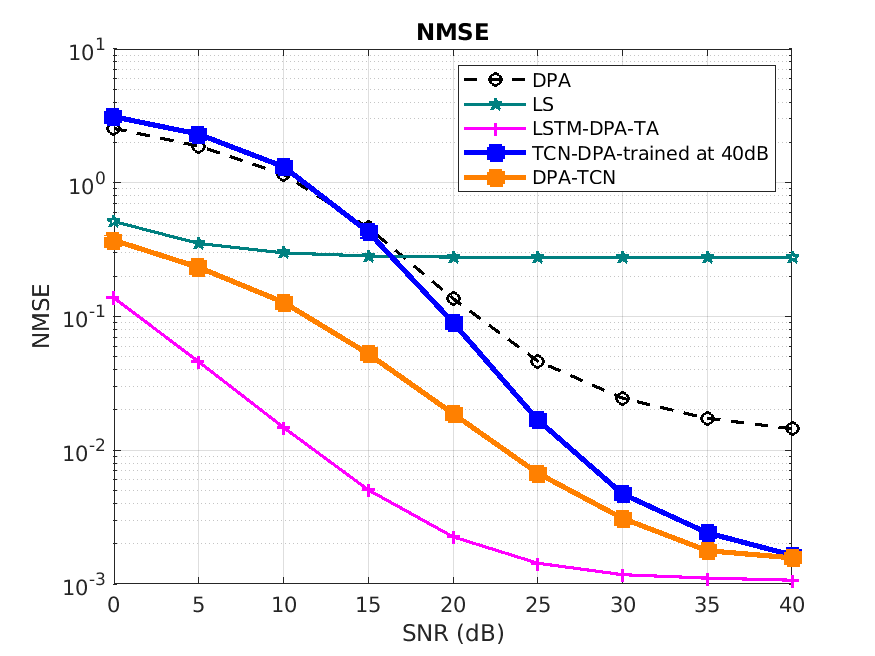} 
        \caption{NMSE performance of various estimators.}
        \label{fig:TCN-w-TA-NMSE}
    \end{subfigure}
    \caption{BER and NMSE simulation results on VTV-SDWW vehicular channel model.}
    \label{fig:with-or-without-TA}
\end{figure}

Fig.~\ref{fig:TCN-no-TA-BER} shows the BER performance of various estimators. 
The LS estimator shows very poor performance across all tested SNRs. A standalone DPA estimator achieves better BER performance than LS but falls short when compared to NN-based estimators, especially at SNRs greater than 20 dB. 
The TCN-DPA trained at 40 dB achieves average performance at low SNRs up to 15 dB, and an exponential improvement from 20 dB to 40 dB. This improved performance in high SNR environments but not in low SNRs indicates the TCN's ability to excel in the conditions in which it is trained, but less so elsewhere. 

Our proposed DPA-TCN estimator, trained using a mixed dataset, demonstrates good performance over a wide SNR range. 
Furthermore, our proposed DPA-TCN estimator achieves an overall BER performance comparable to that of the LSTM-DPA-TA. At SNR levels of 0 to 10 dB, the performance of the two estimators is similar. From 15 dB to 33 dB, the BER using the LSTM-DPA-TA estimator is lower than that when using the TCN estimator. Beyond 33 dB, the TCN estimator performs slightly better than the LSTM-DPA-TA.  

\subsection{NMSE}
The proposed DPA-TCN estimator aims to improve channel estimation by enhancing DPA processing. As illustrated in Fig.~\ref{fig:TCN-w-TA-NMSE}, the DPA-TCN estimator consistently achieves lower NMSE values compared to the standalone DPA estimator and other methods across all SNR ranges investigated. In contrast, the TCN-DPA trained at 40 dB exhibits performance similar to the initial DPA estimator at low SNRs (0 to 15 dB) but shows exponential improvement starting from 20 dB. The LSTM-DPA-TA estimator outperforms all the models presented with significantly lower NMSE values which are only approached by the other presented models at high SNR levels. Our DPA-TCN outperforms the standalone DPA estimator over most of the SNR range, approaching  LSTM-DPA-TA and TCN-DPA performance at very high SNR levels.

\subsection{Complexity analysis}
This section discusses the computational complexity of the estimators presented in this work using real-valued operations as a measure of complexity. Real-valued operations are generally used to analyse the complexity of channel estimation methods~\cite{gizzini2020deep}.  

LS estimation requires the addition and division of the received preambles as shown in (\ref{eqn:ls}), which is \(2X_{A}\) summations and \(2X_{A}\) divisions where $X_A$ represents the number of active subcarriers. In this case, \(X_A = 52\) subcarriers.
From equations (\ref{eqn:yeq}) to (\ref{eqn:fina_DPA}) DPA requires 2 complex value divisions that require 6 real-valued multiplications, 2 real-valued divisions, 2 real-valued summations, and one real-valued subtraction totaling \(26X_{A}\) operations.





In~\cite{gizzini2021temporal}, the authors provide a detailed analysis of the computational complexity of the LSTM-DPA-TA estimator. This analysis includes the real-value operations performed by each LSTM gate and the TA post-processing method. The computational complexity for the LSTM-DPA-TA component is given in~\cite{gizzini2021temporal} as: 
\begin{equation*}
\begin{aligned}
L_\text{LSTM} &= 4(L^{2}) + L(4X_{\text{in}} + 3) + 18X_{D} + X_{\text{in}} \\
& + 13L + 5X_{\text{in}} + 8X_{D} - 8, 
\end{aligned}
\tag{7}
\end{equation*}

where the input size \(X_{\text{in}} = 104\), is obtained by stacking the real and imaginary parts of the complex data along the subcarrier dimension, $X_{D}$ is the input data subcarrier which is also obtained by stacking the real and imaginary parts of the 48 complex data subcarriers, and \(L\) represents the unit size of the LSTM, with \(L = 128\).



The complexity of a TCN residual block depends primarily on its 1D convolutional layers. Our proposed DPA-TCN has a single residual block and performs a one-dimensional convolution along the subcarrier domain. The residual block has 4 layers and uses a kernel size \(K = 2\). Our TCN has the same input  (\(X_I\)) and output (\(X_O\)) sizes, which are equal to the number of active subcarriers \(X_A = 52\). The input data contains 100 features, which are also fed into the TCN as \(X_F\). 

Each convolutional layer requires (\(X_I \times X_F \times K\)) real-valued multiplications and \(X_O\) real-valued summations. The total complexity for 4 layers is calculated as follows:  

\begin{itemize}
    \item Total multiplications:
    \begin{equation*}
        N_{\text{mul}} = 4 \times (X_A \times K \times X_F) = 8(X_A X_F). \tag{8}
        \label{eqn: complex_8}
    \end{equation*}
    
    \item Total summations:
    \begin{equation*}
        N_{\text{sum}} = 4X_A. \tag{9}
        \label{eqn:complex_9}
    \end{equation*}
\end{itemize}
Additionally, considering the residual connection, when the input and output dimensions are different, the multiplications for the residual connection equal \(X_A \times 1\), and the summations for the residual connection are \(X_A\). Thus, the total real-valued multiplications and summations of our TCN estimator are:
%
%
\begin{equation*}
N_{\text{mul,sum}} = 8X_AX_F + 6X_A. \tag{10}
\label{eqn:complex_10}
\end{equation*}
Equations \ref{eqn: complex_8} to \ref{eqn:complex_10} quantify the computational complexity of the TCN estimator in terms of real-valued multiplications and summations. The total complexity is dominated by the convolutional operations, which scale linearly with the number of active subcarriers \(X_A\) and the number of features \(X_F\). 

\begin{table}[htbp]
\centering
\renewcommand{\arraystretch}{1.5} 
\caption{Total computational complexity}
\begin{tabular}{|c|c|}
\hline
\textbf{Estimator} & \textbf{Total complexity} \\
\hline
LS & 208 \\
\hline
DPA & 1,560 \\
\hline
LSTM-DPA-TA & 123,944 \\
\hline
DPA-TCN (proposed) & 43,472 \\
\hline
\end{tabular}
\label{tab:real-valued}
\end{table}

Table~\ref{tab:real-valued} shows the total complexity of the estimators compared in this work. The proposed DPA-TCN estimator has a reduced complexity compared to the LSTM-DPA-TA estimator by approximately 65\%. In other words, the TCN estimator is around 3 times less complex than the LSTM-DPA-TA. Its reduced complexity makes it ideal for vehicular communications that require faster processing. 

\section{Conclusion}
\label{conclusion}
%
%
In this paper, we proposed a novel DPA-TCN estimator for channel estimation in vehicular communications, specifically designed for the VTV-SDWW channel model. By integrating the advantages of DPA and a TCN, the proposed estimator effectively addresses the challenges of channel estimation in high-mobility environments. Simulation results indicate that our proposed DPA-TCN estimator achieves BER performance comparable to the current state-of-the-art estimator (LSTM-DPA-TA) but with significantly reduced computational complexity in comparison. 
Future work will investigate other vehicular channel models.

\section*{Acknowledgment}

The authors thank 
Telkom CoE at the NWU who supported this research.

\bibliographystyle{IEEEtran}

\bibliography{mybibliography}

@inproceedings{gizzini2021temporal,
  title={{Temporal Averaging LSTM-based Channel Estimation Scheme for IEEE 802.11p Standard}},
  author={Gizzini, Abdul Karim and Chafii, Marwa and Ehsanfar, Shahab and Shubair, Raed M},
  booktitle={2021 IEEE Global Communications Conference (GLOBECOM)},
  pages={01--07},
  year={2021},
  organization={IEEE}
}

@article{gizzini2020deep,
  title={{Deep Learning Based Channel Estimation Schemes for {IEEE} 802.11p Standard}},
  author={Gizzini, Abdul Karim and Chafii, Marwa and Nimr, Ahmad and Fettweis, Gerhard},
  journal={IEEE Access},
  volume={8},
  pages={113751--113765},
  year={2020},
  publisher={IEEE}
}

@inproceedings{han2019deep,
  title={{A Deep Learning Based Channel Estimation Scheme for IEEE 802.11p Systems}},
  author={Han, Seungho and Oh, Yeonji and Song, Changick},
  booktitle={2019-2019 IEEE International Conference on Communications (ICC)},
  pages={1--6},
  year={2019}
}

@inproceedings{gizzini2020joint,
  title={{Joint TRFI and Deep Learning for Vehicular Channel Estimation}},
  author={Gizzini, Abdul Karim and Chafii, Marwa and Nimr, Ahmad and Fettweis, Gerhard},
  booktitle={2020 IEEE Global Communications Conference (GLOBECOM) Workshops},
  pages={1--6},
  year={2020},
  organization={IEEE}
}

@article{pan2021channel,
  title={{Channel Estimation Based on Deep Learning in Vehicle-to-Everything Environments}},
  author={Pan, Jing and Shan, Hangguan and Li, Rongpeng and Wu, Yingxiao and Wu, Weihua and Quek, Tony QS},
  journal={IEEE Communications Letters},
  volume={25},
  number={6},
  pages={1891--1895},
  year={2021},
  publisher={IEEE}
}

@inproceedings{akiba2019optuna,
  title={{Optuna: A Next-generation Hyperparameter Optimization Framework}},
  author={Akiba, Takuya and Sano, Shotaro and Yanase, Toshihiko and Ohta, Takeru and Koyama, Masanori},
  booktitle={Proceedings of the 25th ACM SIGKDD International Conference on Knowledge Discovery \& Data mining},
  pages={2623--2631},
  year={2019}
}

@article{lecun1995convolutional,
  title={{Convolutional Networks for Images, Speech, and Time-Series}},
  author={LeCun, Yann and Bengio, Yoshua and others},
  journal={The handbook of brain theory and neural networks},
  volume={3361},
  number={10},
  pages={1995},
  year={1995},
  publisher={Citeseer}
}

@article{bai2018empirical,
  title={{An Empirical Evaluation of Generic Convolutional and Recurrent Networks for Sequence Modeling}},
  author={Bai, Shaojie and Kolter, J Zico and Koltun, Vladlen},
  journal={arXiv preprint arXiv:1803.01271},
  year={2018}
}

@INPROCEEDINGS{7421323,
  author={Walk, Philipp and Becker, Henning and Jung, Peter},
  booktitle={2015 49th Asilomar Conference on Signals, Systems and Computers}, 
  title={{OFDM Channel Estimation via Phase Retrieval}}, 
  year={2015},
  volume={},
  number={},
  pages={1161-1168},
  doi={10.1109/ACSSC.2015.7421323}}

@INPROCEEDINGS{10279602,
  author={Jovane, Juan D. and Lee, Chia-Han},
  booktitle={ICC 2023 - IEEE International Conference on Communications}, 
  title={{Channel Estimation using Temporal Convolutional Networks for V2X Communications}}, 
  year={2023},
  volume={},
  number={},
  pages={565-570},
  doi={10.1109/ICC45041.2023.10279602}}

@INPROCEEDINGS{4526014,
  author={Jiang, Daniel and Delgrossi, Luca},
  booktitle={2008 VTC Spring - IEEE Vehicular Technology Conference}, 
  title={{{IEEE} 802.11p: Towards an International Standard for Wireless Access in Vehicular Environments}}, 
  year={2008},
  volume={},
  number={},
  pages={2036-2040},
  doi={10.1109/VETECS.2008.458}}

@INPROCEEDINGS{4350097,
  author={Acosta-Marum, Guillermo and Ingram, Mary Ann},
  booktitle={2007 IEEE 66th Vehicular Technology Conference}, 
  title={{Six Time- and Frequency-Selective Empirical Channel Models for Vehicular Wireless LANs}}, 
  year={2007},
  volume={},
  number={},
  pages={2134-2138},
  doi={10.1109/VETECF.2007.448}}

@inproceedings{Ngorima2024,
  author={Simbarashe Aldrin Ngorima and Albert Helberg and Marelie H. Davel},
  title={{A Data Pilot-Aided Temporal Convolutional Network for Channel Estimation in IEEE 802.11p Vehicle-to-Vehicle Communications}},
  booktitle={Southern Africa Telecommunication Networks and Applications Conference (SATNAC)},
  year={2024}
}

@InProceedings{ngorimaSA,
author={Simbarashe Aldrin Ngorima and Albert Helberg and Marelie H. Davel},
editor="Pillay, Anban
and Jembere, Edgar
and J. Gerber, Aurona",
title={{Sequence Based Deep Neural Networks for Channel Estimation in Vehicular Communication Systems}},
booktitle="Artificial Intelligence Research",
year="2023",
publisher="Springer Nature Switzerland",
address="Cham",
pages="172--186",
isbn="978-3-031-49002-6"
}

@InProceedings{NgorimaSACAIR2024,
author="Ngorima, Simbarashe Aldrin
and Helberg, Albert S. J.
and Davel, Marelie H.",
editor="Gerber, Aurona
and Maritz, Jacques
and Pillay, Anban W.",
title={{Neural Network-Based Vehicular Channel Estimation Performance: Effect of Noise in the Training Set}},
booktitle="Artificial Intelligence Research",
year="2025",
publisher="Springer Nature Switzerland",
address="Cham",
pages="192--206"
}

\end{document}